\documentclass[pdflatex,sn-basic]{sn-jnl}

\usepackage{graphicx}%
\usepackage{multirow}%
\usepackage{amsmath,amssymb,amsfonts}%
\usepackage{amsthm}%
\usepackage{mathrsfs}%
\usepackage[title]{appendix}%
\usepackage[dvipsnames]{xcolor}%
\usepackage{textcomp}%
\usepackage{manyfoot}%
\usepackage{booktabs}%
\usepackage{algorithm}%
\usepackage{algorithmicx}%
\usepackage{algpseudocode}%
\usepackage{listings}%
\usepackage{graphicx}
\usepackage{txfonts}
\usepackage{lipsum}
\usepackage{natbib}
\usepackage{ulem}
\usepackage{lineno}

\theoremstyle{thmstyleone}%
\theoremstyle{thmstyletwo}%

\theoremstyle{thmstylethree}%

\definecolor{bouliches}{rgb}{0, 115, 158}

\begin{document}

\title[Quantum modified inertia]{Quantum modified inertia: Application to galaxy rotation curves}


\author*[1]{\fnm{Jonathan} \sur{Gillot}}\email{jonathan.gillot@femto-st.fr}

\affil*[1]{\orgdiv{Universit\'e Marie et Louis Pasteur}, \orgname{Supmicrotech, FEMTO-ST,
		Observatoire des Sciences de l'Univers THETA}, \orgaddress{\street{26 rue de l'Epitaphe}, \city{Besan\c con}, \postcode{25000}, \country{ France}}}



\abstract{This work explores modified inertia in the context of galactic dynamics by investigating the consequences of introducing bounds on acceleration.
Building on earlier ideas related to maximal acceleration and quantum speed limits, an effective framework is developed in which both upper and lower acceleration bounds are incorporated into the formalism of special relativity, motivated by a conjectured correspondence between the proper time of an accelerated object and the quantum speed limit.

A characteristic minimal acceleration of order  $1.8 \times 10^{-11}$ m s$^{-2}$ naturally emerges from the analysis. The resulting modified inertia model is applied to galaxy rotation curves, taking into account the baryonic contributions from stellar disks, gas, and bulges. An analytic expression for the radial acceleration relation is derived within this framework. The resulting radial acceleration relation is consistent with the Cassini quadrupole constraint in the Solar System, and remains compatible with recent observational results on dwarf spheroidals and ultra-wide binaries. When confronted with observations, the model provides a good description of the Milky Way and the dwarf galaxy DDO 52 rotation curves. It also successfully recovers the Tully-Fisher relation linking the baryonic mass and the terminal velocity, with $\mathrm{log}(M) = 4 \, \mathrm{log}(v) + 2.62$.

Within the proposed model, the presence of a lower acceleration bound significantly reduces the amount of dark matter required to account for galaxy rotation curves. Possible implications for the redshift evolution of this minimal acceleration are briefly discussed.}




\maketitle

\section{Introduction}\label{sec1}

The long-standing problem of galaxy rotation curves \cite{Zwicky09,bosma_distribution_1978,Rubin70,Rubin80,deSwart17} has motivated the introduction of a new form of matter, commonly referred to as dark matter. On cosmological scales, this component is inferred to exceed the baryonic matter content by roughly a factor of five and is invoked to explain galactic rotation curves, galaxy dynamics within clusters \cite{Mcmillan11}, and large-scale structure formation \cite{krishnan_clustering_2020}. A wide range of observations, including X-ray measurements of galaxy clusters \cite{bradac_dark_2008}, gravitational lensing \cite{umetsu_clustergalaxy_2020}, and cosmic microwave background anisotropies \cite{chacko_hidden_2015}, have been used to constrain the properties of dark matter. Numerous candidates spanning more than ninety orders of magnitude in mass have been proposed, from compact objects \cite{Yoo04} to ultra-light scalar fields \cite{Stadnik15}, WIMPs \cite{Roszkowski18}, axions \cite{ADMX18,Savalle21}, and sterile neutrinos \cite{Boyarsky19}. Despite extensive experimental efforts \cite{Gaskins16,Bertone18}, no non-gravitational detection of dark matter has yet been confirmed.

In parallel, alternative approaches have been developed in which dark matter is absent or significantly reduced. Among them, Modified Newtonian Dynamics (MOND) \cite{milgrom_modification_1983-1, Milgrom08} postulates a modification of dynamics or gravity in the low-acceleration regime, introducing a characteristic acceleration scale $a_0 \simeq 1.2\times10^{-10}\,\mathrm{m\,s^{-2}}$. This scale, empirically inferred from galaxy rotation curves, has been shown to reproduce a wide range of galactic kinematics \cite{lelli_one_2017}. Early on, Milgrom suggested a possible connection between $a_0$ and cosmological parameters \cite{Milgrom94}, an idea reinforced after the discovery of cosmic acceleration \cite{Riess98}, leading to relations of the form $a_0 \sim a_\Lambda = c^2\sqrt{\Lambda/3}$ \cite{milgrom_a_0_2020}. Related ideas also arise in quantized inertia models, which predict a minimal acceleration linked to the size of the observable Universe \cite{Mcculloch17}.

Independently, the concept of a fundamental maximal acceleration has been explored within quantum and relativistic frameworks. Caianiello first introduced the notion of a maximal acceleration \cite{Caianiello81}, later refined and extended in various contexts \cite{Wood89,Pati92,Pati92a,Papini02,termini_imagination_2006}. More recently, Rovelli and Vidotto argued that such a bound is required to avoid curvature singularities in quantum gravity \cite{Rovelli13}. Given the prevalence of symmetry principles in physics, the existence of a maximal acceleration naturally raises the question of a corresponding minimal acceleration bound.

In this work, we investigate the theoretical foundations of modified inertia by introducing bounds on proper time within the framework of special relativity (SR). These bounds, which are conjecturally associated with quantum speed limits (QSL), naturally lead to effective maximal and minimal acceleration bounds. Owing to the extremely small expected magnitude of the minimal acceleration $a_{\rm min}$, galactic dynamics provides a particularly relevant observational testing ground.
		
Although the phenomenology explored here operates in a low-acceleration regime reminiscent of MOND, the present approach is conceptually different. In particular, the acceleration limit introduced in this work is not a transition parameter interpolating between two dynamical regimes, but rather a lower acceleration bound motivated by quantum and relativistic considerations, without the introduction of empirical interpolation functions.

We first present the construction of the theory, derive the corresponding dynamical relation and propose a conjecture on the nature of the acceleration bounds. We then investigate the implications of the minimal acceleration for galaxy rotation curves using two different morphological systems (DDO~52 and the Milky Way), as well as for the radial acceleration relation (RAR), wide binaries at large separations, and the baryonic Tully-Fisher relation (TFR). Finally, we discuss the domain of validity of this SR-extended modified dynamics and briefly explore its possible cosmological implications.



\section{Fundamental dynamics principles reshaped by quantum bounds}
SR is suitable for treating accelerations produced by any force, except for strong gravitational regimes. For the moment, we consider a force in the general sense of the definition without specifying its nature, and we try to derive a general expression of the principle of dynamics. Acceleration in SR has been extensively studied, notably by Rindler who described the motion of a uniformly accelerated rod \cite{rindler_kruskal_1966}.\\
\indent In SR, the proper acceleration $\alpha$ is the acceleration experienced by an object of mass $m$ and measurable with an accelerometer, which measures the quantity $\boldsymbol{\alpha}=\mathbf{F}/m$, where $\mathbf{F}$ is the proper force. Thus, proper acceleration is relative to an inertial observer who is instantly and momentarily at rest relative to the accelerated object. The proper acceleration is different from the coordinate acceleration $\textbf{a}$ which is dependent on the chosen coordinate system. 
For arbitrary direction of the velocity $\boldsymbol{v}$, the three-coordinate acceleration $\boldsymbol{a}$ can be expressed as a function of three-proper force $\boldsymbol{F}$ and the Lorentz factor $\gamma=(1-v^2/c^2)^{-1/2}$:

\begin{equation}
	\boldsymbol{a}=\frac{1}{m \gamma}\left(\boldsymbol{F} - \frac{(\boldsymbol{F \cdot v})\boldsymbol{v}}{c^2} \right)
\end{equation}

If we consider that the velocity is parallel to the $x$ axis, the coordinate acceleration along each direction is written 

\begin{eqnarray} \label{three-acc}
	&\boldsymbol{a_x}&=\frac{1}{\gamma^3} \frac{F_x}{m}  \nonumber \\
	&\boldsymbol{a_y}&=\frac{1}{\gamma} \frac{F_y}{m}  \\
	&\boldsymbol{a_z}&=\frac{1}{\gamma} \frac{F_z}{m}. \nonumber \\
\nonumber
\end{eqnarray}

For the following steps of the demonstration, we will consider the case of a force applied in the direction of the velocity. The proper acceleration $\alpha$ for uni-directional constant acceleration can also be expressed using the rapidity $\boldsymbol{\eta}$ 

\begin{equation} \label{properacc2}
	\boldsymbol{\alpha}=c \frac{d\boldsymbol{\eta}}{d\tau}=\gamma^3 \frac{d\boldsymbol{v}}{dt}=\gamma^3 \boldsymbol{a}.
\end{equation}

Moreover, rapidity is defined too as

\begin{equation} \label{rapido}
	\boldsymbol{\eta}= \text{atanh} \left(\frac{{v}}{c}\right) \boldsymbol{\hat{v}},
\end{equation}

where $\boldsymbol{\hat{v}}=\boldsymbol{v}/|v|$. The integration over $\tau$ of the first expression of (\ref{properacc2}) leads to

\begin{equation} \label{rapido2}
	\boldsymbol{\eta}= \frac{\boldsymbol{\alpha} \tau}{c},
\end{equation}

and injecting (\ref{rapido2}) in (\ref{rapido}) gives

\begin{equation} \label{vfctdet}
	\boldsymbol{v}=c\ \text{tanh}\left(\frac{\alpha \tau}{c}\right) {\hat{\boldsymbol{\alpha}}},
\end{equation}

with $\hat{\boldsymbol{\alpha}}=\boldsymbol{\alpha}/|\alpha|$. Now that we have an expression for the coordinate velocity $v$, we can obtain the coordinate acceleration $\boldsymbol{a}$ with the time derivative with respect to $t$. However, the right-hand term has no explicit dependence in $t$, and $v$ has no explicit dependence on $\tau$. We can yet write:

\begin{equation} \label{aetape1}
	\frac{d}{dt} \left( \frac{dt}{d \tau} \boldsymbol{v} \right)= \frac{d}{d \tau} c\  \text{tanh}\left(\frac{{\alpha} \tau}{c}\right) {\hat{\boldsymbol{\alpha}}}.
\end{equation}

By definition, $\gamma={dt}/{d \tau}$ and with an algebraic development ${d (\gamma v)=\gamma^3 dv}$. The dependence on $\gamma^3$ can change if the acceleration is not parallel to velocity. However, for the longitudinal acceleration we obtain 

\begin{equation} \label{aetape2}
	\gamma^3 \boldsymbol{a}= c \frac{d}{d \tau} \text{tanh}\left(\frac{{\alpha} \tau}{c}\right) {\hat{\boldsymbol{\alpha}}}.
\end{equation}

At this stage, the main novelty of the derivation is the assumption that the SR proper time $\tau$ cannot become arbitrarily small or arbitrarily large, but is instead bounded by two extremal values, denoted $\tau^{+}$ and $\tau^{-}$. Within this interpretation, $\tau^{+}$ and $\tau^{-}$ respectively correspond to the minimal and maximal evolution times associated with the dynamical evolution of a physical system. Their possible physical interpretation, in connection with quantum speed limits and causal constraints, will be discussed in the next subsection.

\indent The proposed hypothesis imposes that we cannot consider the formal definition of the derivative

\begin{equation} \label{derivmath}
	\frac{d f}{d \tau } \bigg|_{\tau=\tau_0} =\lim_{\Delta \tau \to 0\atop \Delta \tau \ne0} \frac{f(\tau_0+\Delta \tau)-f(\tau_0)}{\Delta \tau},
\end{equation}

In this framework, we assume that at the initial time, the observer and the object share the same inertial reference frame. The object then undergoes a constant proper acceleration during a time interval $\tau$. At the end of this interval, the object has acquired a new instantaneous velocity. At this stage, knowledge of this velocity alone is sufficient to determine the corresponding time dilation, in accordance with the clock hypothesis. If the object subsequently experiences a new acceleration episode, not necessarily identical to the previous one, the situation is formally identical to the initial configuration: the observer can again be considered momentarily at rest in the instantaneous rest frame of the object at the onset of this new acceleration. This acceleration remains constant over a duration $\tau$, producing an additional velocity increment. Repeating this process, the motion can be described as a sequence of acceleration steps.


The equation (\ref{aetape2}) is consequently rewritten as a set of two equations :

\begin{equation} \label{amaxcoef}
	\boldsymbol{a'} = \frac{1}{\gamma^3} \left\{ \kappa \frac{c}{ \tau^+} \text{tanh}\left(\frac{{\alpha} \tau^+}{c}\right) + \epsilon \right\} \hat{\boldsymbol{\alpha}}
\end{equation} 
\begin{equation} \label{amincoef}	
	\boldsymbol{a''} = \frac{1}{\gamma^3} \left\{ \beta \frac{c}{ \tau^-} \text{tanh}\left(\frac{{\alpha} \tau^-}{c}\right) + \chi \right\} {\hat{\boldsymbol{\alpha}}},
\end{equation}

with $\kappa, \epsilon, \beta$ and $\chi$ parameters that have to be determined later. Since these parameters are expected to be scalar equal to 0 or $\pm$ 1, they can be arbitrarily added. Because it is assumed that $\tau^{+}$ and $\tau^{-}$ do not tend respectively  to $0$ and $+\infty$, the consequence is to introduce two bounds on dynamics homogeneous to accelerations:
	
\begin{equation}
a_{\rm max}	\triangleq \frac{c}{ \tau^+},
\end{equation}
\begin{equation}	
a_{\rm min} \triangleq	\frac{c}{ \tau^-}.
\end{equation}


\indent We now examine the boundary conditions that reconcile these two competing effects. The combined contributions of maximal and minimal acceleration are expected to recover the Newtonian law of dynamics over a wide range of accelerations, with deviations occurring only in the ultra-low and ultra-high acceleration regimes. Accordingly, we propose to write the dynamical equation as the sum of these two effects, following the form introduced in equations (\ref{amaxcoef}) and (\ref{amincoef}):
\begin{eqnarray} \label{a+1coef}
\boldsymbol{a} &=&\boldsymbol{a'}+\boldsymbol{a''}\nonumber \\
\boldsymbol{a} &=& \frac{1}{\gamma^3} \left\{ \kappa  a_{\rm max} \text{tanh}\left(\frac{\alpha}{a_{\rm max}}\right) + \epsilon + \beta a_{\rm min} \text{tanh}\left(\frac{\alpha}{a_{\rm min}}\right) + \chi \right\} \hat{\boldsymbol{\alpha}}
\end{eqnarray}

The first condition is the limit $\boldsymbol{\alpha} =  0$ imposing that ${\boldsymbol{a} = a_{\rm min}/\gamma^3} \hat{\boldsymbol{\alpha}}$. This minimal acceleration limit leads to
$(\epsilon+\chi)/\gamma^3=a_{\rm min}$, because the terms in tanh of (\ref{a+1coef}) vanish.

The second condition is given by the maximal boundary condition ${\boldsymbol{\alpha} = \pm \infty \hat{\boldsymbol{\alpha}}}$, for which we should have ${\boldsymbol{a} = \pm a_{\rm max}/\gamma^3} \hat{\boldsymbol{\alpha}}$, and not ${\boldsymbol{a} = (a_{\rm max}+a_{\rm min})/\gamma^3} \hat{\boldsymbol{\alpha}}$, otherwise it would violate the maximal permitted acceleration. Thus, it follows that $\kappa=+1$ and that $\beta = -1$.
Thus, after replacing the proper acceleration $\alpha$ by the proper force $F$ previously introduced, the equation governing the dynamics limited by temporal bounds can be written as follows:

\begin{equation} \label{PFD}
	\boldsymbol{a} = \frac{1}{\gamma^3} \left\{ a_{\rm max}\ \text{tanh}\left(\frac {F}{m a_{\rm max}}\right) + a_{\rm min} \left[ 1 -  \text{tanh}\left(\frac {F}{m a_{\rm min}}\right)  \right]\right\} {\hat{\boldsymbol{F}}},
\end{equation}\\
where $\hat{\boldsymbol{F}}=\mathbf{F}/|F|$.

\begin{figure}
	
	\center
	\includegraphics[width=9cm]{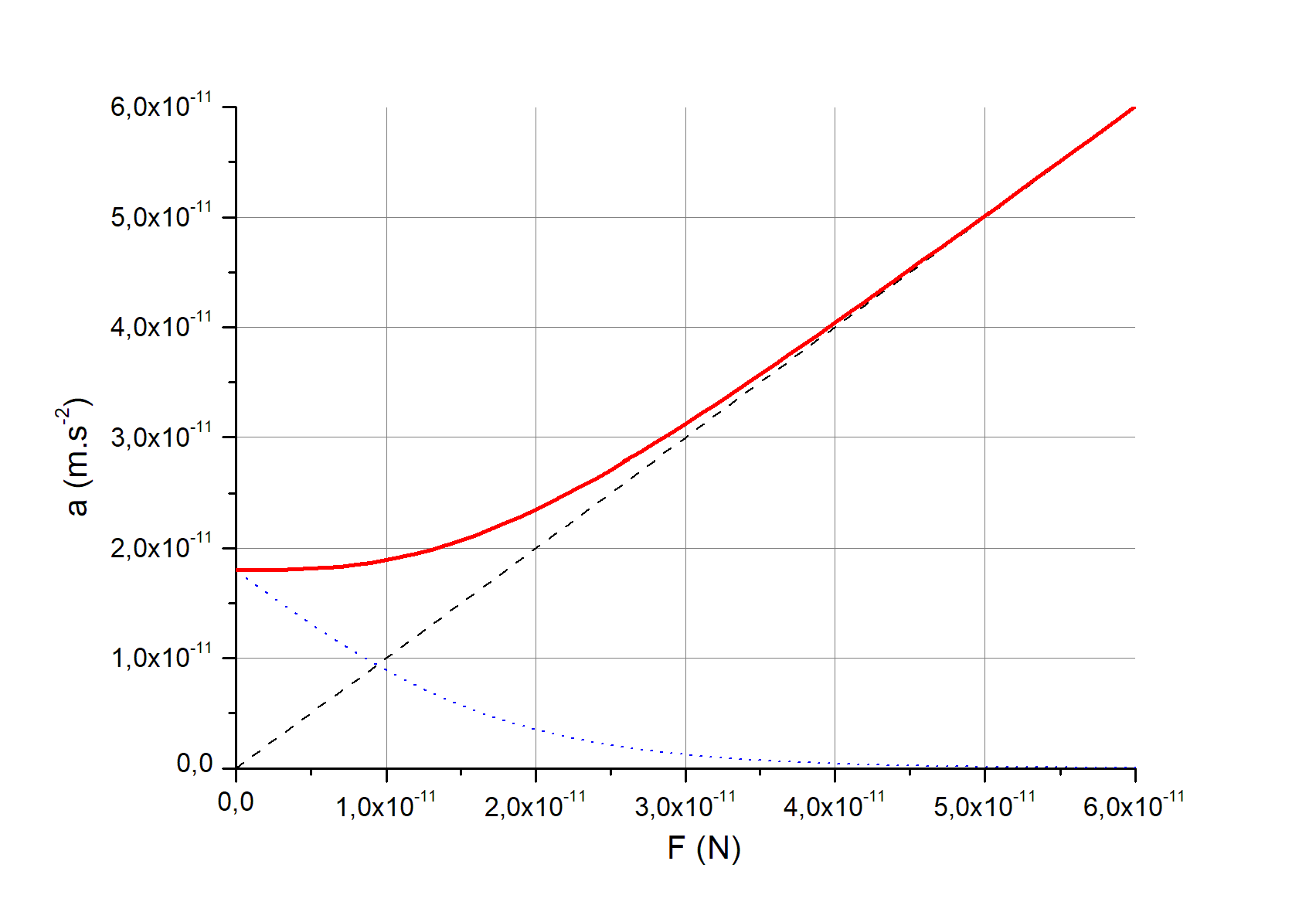}
	
	\caption{Plot of the two different terms of equation (\ref{PFD}) for $m=1$ and $a_{\rm min} = 1.8 \times 10^{-11}$ m s$^{-2}$. The black dashed line is the plot of first term associated to the maximal acceleration bound. The blue dotted line is the plot of the second term assigned to minimal acceleration limit. The continuous red line is the equation (\ref{PFD}). The acceleration saturation effect at maximal acceleration is symmetric and not represented.}
	\label{accel-decort}
\end{figure}

The equation (\ref{PFD}) is plotted in figure \ref{accel-decort}, with a zoom near $F=0$, close to the minimal acceleration bound. Within this dynamical framework, the relation between force and acceleration becomes intrinsically nonlinear, departing from the standard proportionality of Newtonian and relativistic dynamics. Although the construction is based on special-relativistic kinematics, the resulting dynamics no longer corresponds to the usual local differential description of motion. Instead, the acceleration is effectively constrained by finite evolution times. In this sense, QMI may be interpreted as an effective extension of special relativity, in which standard relativistic dynamics is smoothly recovered in the limits $\tau^+_{QSL}\rightarrow 0$ and $\tau^-_{QSL}\rightarrow \infty$. With these combinations of two hyperbolic tangents, we identify three acceleration regimes:

\begin{itemize}
	
	\item The first regime is Newtonian and appears if ${ma_{\rm min}\ll F \ll m  a_{\rm max}}$. In this case, the principle of dynamics is almost identical to that of Newton, and the dynamics are well approximated by $a=F/m$. Because the minimal acceleration term vanishes, the proportionality relation $a=F/m$ is ensured by the hyperbolic tangent function properties, since ${\text{tanh}(x)=x}$ when ${x \rightarrow 0}$, recovering the classical limit if $v<<c$.
	
	\item The second regime involves an applied force that is close to the maximal acceleration bound. The dynamics enter a maximal acceleration regime, where ${F \sim m \, a_{\rm max}}$. Newton's law no longer applies and the acceleration of the body becomes less and less strong even if the applied force increases indefinitely. At some point, the acceleration is locked at the value $a_{\rm max}$ even if ${F \rightarrow \infty}$.
	
	\item The third regime, more described in this paper, appears when $F \sim m \, a_{\rm min}$. Symmetrically to the maximal regime, the acceleration is no longer propor\-tional to the applied force, but begins to smoothly saturate at a value $a_{\rm min}$ as $F \rightarrow 0$. In figure \ref{accel-decort}, we can see that when $F$ is decreasing, the minimal acceleration term plotted in the dashed blue line becomes dominant whereas the maximal acceleration term in the dotted black line tends to 0 quasi linearly, because this is the maximal term that approximates the typical Newtonian relation $a=F/m$.
	
\end{itemize} 

The next section briefly reviews theoretical bounds associated with the maximal and minimal rates of temporal evolution of physical systems, with particular emphasis on quantum speed limits (QSL). This discussion motivates the conjectured correspondence between QSL-related evolution times and the proper timescales $\tau^+$ and $\tau^-$ introduced in the present framework.

\subsection{Maximal accelerations in modern physics}

The notion of a maximal proper acceleration has emerged in several theoretical contexts at the intersection of quantum mechanics, relativity, and gravity. The earliest systematic formulation was proposed by Caianiello \cite{Caianiello81, caianiello_maximal_1984-1}, who argued that the Heisenberg uncertainty principle, when applied to relativistic kinematics, implies a fundamental upper bound on the proper acceleration of a point particle. By introducing a geometric structure on phase space and associating quantum transitions with a minimal proper time, Caianiello derived a mass-dependent maximal acceleration

\begin{equation}
	a_{\rm max} \sim \frac{2 m c^3}{\hbar},
\end{equation}

which scales linearly with rest mass energy. In this framework, maximal acceleration appears as a kinematical bound imposed by quantum principles rather than by dynamics. These ideas were later refined by Papini and Feoli \cite{termini_imagination_2006, feoli_maximal_2003}, who connected the bound to the QSL governing the evolution of orthogonal states and explored its phenomenological implications. 

The concept was subsequently applied to extended objects by Frolov and S\'anchez \cite{frolov_instability_1991}, who investigated uniformly accelerated open relativistic strings in flat spacetime. They demonstrated the existence of a critical acceleration \(a_c\) above which rigid equilibrium configurations cease to exist. This instability arises from the causal structure of spacetime: at high acceleration, the Rindler horizon  overlap the string endpoints, preventing causal communication along the string. A closely related result was obtained by Gasperini \cite{gasperini_kinematic_1991}, who showed that open strings in a gravitational background become unstable when the proper acceleration exceeds a critical value, again due to the loss of causal contact between different parts of the string. In contrast to point-particle formulations, these limits emerge dynamically from the finite size and internal structure of extended objects.

More recently, Rovelli and Vidotto \cite{Rovelli13} argued that a maximal proper acceleration arises naturally in loop quantum gravity. In their approach, the discreteness of quantum geometry implies an upper bound on measurable acceleration, independent of particle mass or spatial extension, and rooted in the minimal area and volume quanta of spacetime. 

Taken together, these works trace a coherent evolution of the concept of maximal acceleration. Despite their different physical motivations, these studies underscore that maximal acceleration is not merely a formal curiosity but a recurring and potentially universal feature.

The QSL establishes a fundamental bound on the minimum time $\tau^+_{QSL}$ required for a quantum system to evolve between two distinguishable states \cite{mandelstam_uncertainty_1945, margolus_maximum_1998} (see discussion about QSL in \ref{appendixA}). As shown in the previous section, this temporal bound can be translated into a constraint on the rate of change of the velocity when combined with relativistic kinematics, effectively defining a maximal proper acceleration \cite{papini_revisiting_2003,feoli_maximal_2003}. In this interpretation, maximal acceleration arises as a quantum-kinematical limit rooted in the energy-time uncertainty principle.

Here, we adopt a first assumption that the QSL, which is fundamentally tied to the rest energy of the system, can be related to the proper time experienced by an accelerated particle in relativity. This correspondence implies that the kinetic energy of an object cannot vary arbitrarily fast, but is bounded by the quantum limit on state evolution. Once we substitute $\tau_{QSL}^+={\pi \hbar}/{2mc^2}$ in the equation (\ref{aetape2}), we get 
\begin{equation} \label{aetape3}
	\boldsymbol{a'}= \frac{1}{\gamma^3} \frac{2mc^3}{\pi \hbar} \text{tanh}\left(\frac{\pi \hbar {F}}{2m^2c^3}\right) {\hat{\boldsymbol{F}}}.
\end{equation}
The prefactor $({2mc^3/\pi \hbar})$ corresponds to the Caianiello maximal acceleration \cite{caianiello_maximal_1984-1}. The coordinate acceleration therefore naturally incorporates an upper quantum bound once the minimal quantum evolution time is identified with the relativistic proper time. The definition of $\tau_{QSL}^+$ may depend on the nature of the underlying system, which can be either quantum or classical. Although it is sometimes argued that quantum speed limits are not directly applicable to classical dynamics, several works have shown that classical systems can also exhibit finite speed limits in the limit $\hbar \rightarrow 0$ \cite{shanahan_quantum_2018, garcia-pintos_unifying_2022}. The relation between quantum speed limits and classical speed limits (CSL) is subtle and has been extensively discussed in the literature \cite{bolonek-lason_classical_2021}. In the present work, we do not specify the origin of $\tau^+$ and treat it as an effective minimal evolution timescale entering the relativistic dynamics. In the applications considered below, we focus on regimes governed by the minimal acceleration, so that the quantum or classical interpretation of $\tau^+$ does not affect the phenomenological results.

\subsection{Concepts of minimal acceleration}

Beyond the concept of maximal acceleration, several theoretical studies have explored the possibility of a minimal acceleration, corresponding to a lower bound on the acceleration experienced by a physical system. Such a notion has primarily emerged in cosmological contexts, modified inertia frameworks and quantum gravity considerations. A prominent example of modified inertia is MOND \cite{milgrom_modification_1983-1}, which introduces a characteristic acceleration scale \(a_0 \sim 10^{-10}\,\mathrm{m\,s^{-2}}\) below which departures from Newtonian dynamics occur. Although \(a_0\) does not represent a minimal acceleration, it
signals the transition between local Newtonian behavior and a regime potentially influenced by cosmological effects.\\
\indent From a quantum perspective, the structure of the vacuum may also suggest the existence of a minimal acceleration. Unruh \citet{unruh_notes_1976} showed that a uniformly accelerated observer perceives a thermal bath with temperature \(T = \hbar a/(2\pi c k_B)\), implying that arbitrarily small accelerations correspond to arbitrarily low Unruh temperatures. The existence of a minimal acceleration would therefore imply a lowest observable Unruh temperature, possibly linked to cosmological parameters. In this spirit, McCulloch's quantized inertia hypothesis \citet{Mcculloch17} relates inertia to horizon-scale physics, predicting a minimal acceleration of order \(cH_0\), where \(H_0\) is the Hubble parameter. In addition, Grumiller et Preis derive an effective gravitational model at large distances in which the usual Newtonian potential is supplemented by a constant Rindler acceleration term. This additional "Rindler force" can modify test-particle dynamics over galactic scales and has been suggested to influence rotation curves and anomalies like the Pioneer acceleration. While the model does not predict a fundamental minimal acceleration, it introduces a constant acceleration term that effectively alters the low-acceleration regime of gravity at large distances \cite{grumiller_rindler_2011, lin_galaxy_2013}.\\
\indent Symmetrically to the usual QSL time $\tau^+_{QSL}$, one may consider the possibility of a non-infinite characteristic timescale $\tau^-_{QSL}$ constraining how slowly a system can evolve. Related concepts have recently been discussed in the context of reverse quantum speed limits (RQSL) \cite{mohan_reverse_2021, zhu_observation_2025}. In contrast to the standard QSL, however, the physical interpretation and general formulation of such lower evolution bounds remain less explored. The expression of $\tau^-_{QSL}$ introduced here should therefore be regarded as a  conjecture rather than as a rigorous derivation. The reduced Compton wavelength is defined as $\lambdabar_{\rm C}=\hbar/mc$, allowing the standard QSL timescale to be rewritten as

\begin{equation}
\tau^+_{QSL}=\frac{\pi \hbar}{2mc^2}
=\frac{\pi}{2c}\lambdabar_{\rm C}.
\end{equation}

In this form, $\tau^+_{QSL}$ can be interpreted as being of the order of the time required for a signal to traverse the reduced Compton wavelength.
Motivated by the idea that a quantum object cannot be treated as strictly point-like, we assume that its effective spatial extension is of the order of $\lambdabar_{\rm C}$. In the spirit of the work of Frolov and S\'anchez \cite{frolov_instability_1991}, the associated maximal acceleration could then be interpreted as a causal limitation induced by the Rindler horizon associated with acceleration. More precisely, an object whose spatial extent is comparable to its Compton wavelength may lose internal causal coherence when an acceleration-induced horizon separates regions spanning at least one wavelength. This provides a physically motivated criterion under which causal disconnection could arise for systems subjected to sufficiently large accelerations.\\
\indent By analogy, one may conjecture that the largest admissible wavelength should remain bounded by the diameter of the causally connected region, here identified with a cosmological horizon scale $2R_u$, subsequently referred to as the causality sphere. Preserving the same prefactor $\pi/2c$ then leads to the conjectured expression:

\begin{equation}\label{conjecturette-tau-}
\tau^-_{QSL}=\frac{\pi R_u}{c},
\end{equation}
corresponding to a time of the order of the light-crossing time of the observable causal region.\\

Within this framework, a minimal acceleration may be interpreted as the smallest admissible rate of change of velocity, complementing the maximal-acceleration bounds proposed by Caianiello \cite{Caianiello81}. Although considerably less studied than the maximal-acceleration sector, this approach suggests that dynamics is bounded both at high and low accelerations. With both acceleration bounds incorporated into a unified dynamical expression (Eq. \ref{PFD}), we can now investigate the phenomenological implications of this non-Newtonian dynamics. In the following, we focus on the minimal-acceleration sector and on the transition between this regime and the standard Newtonian limit.

\section{Application to galaxy rotation curves}\label{sec2}

\subsection{Minimal acceleration fiduciary value}
\indent Low-acceleration regimes are difficult to observe in the usual laboratory experiments, even for the best quantum accelerometers \cite{geiger_high-accuracy_2020,leveque_gravity_2021}. Some potential instruments could however reach a sensitivity below $10^{-10}\ \text{m s}^{-2}$ \cite{leveque_carioqa_2023} in a near future.\\
\indent However, this acceleration magnitude is commonly observed in astrophysical structures, since they are often weakly gravitationnally bound. In particular, galaxies provide a natural laboratory.
For that purpose, we can develop a preliminar simple model. We limit the study to objects with circular orbits and weak gravitational regime. The gravitation is thus described by the Newtonian law $\textbf{F}=-GMm\hat{\boldsymbol{r}}/r^2$. The dynamics law (\ref{PFD}) then becomes:

\begin{equation} \label{PFDgrav}
	\boldsymbol{a} = \frac{1}{\gamma^3} \left\{ a_{\rm max}\ \text{tanh}\left(\frac {GM}{ a_{\rm max} r^2}\right) + a_{\rm min} \left[ 1 -  \text{tanh}\left(\frac {GM}{a_{\rm min}r^2} \right)  \right]\right\} {\hat{\boldsymbol{F}}}.
\end{equation}

The circular velocity is given by a simple application of the equation $a=v^2/r$, and the power of $\gamma$ changes because the acceleration and the velocity vectors are perdendicular during the circular motion. In the low acceleration regime (weak gravitational field), the hyperbolic tangent reduces to
\begin{equation} 
	\text{tanh}\left(\frac{GM}{a_{\rm max}r^2}\right)\rightarrow \frac{GM}{a_{\rm max}r^2} ~\text{when}~  \left|\frac{GM}{a_{\rm max}r^2}\right|\rightarrow~0.
\end{equation}
For non-relativistic velocities ($v\ll c$), we get:

\begin{equation} \label{veloce3}
	\boldsymbol{v}=\sqrt{ \frac{GM}{r} + r a_{\rm min} \left[ 1 -  \text{tanh}\left(\frac{GM}{a_{\rm min} r^2}\right)  \right]  } \ \hat{\boldsymbol{r}},
\end{equation}
with  $\hat{\boldsymbol{r}}=\boldsymbol{r}/|r|$.

\begin{figure}
	
	\center
	\includegraphics[width=9cm]{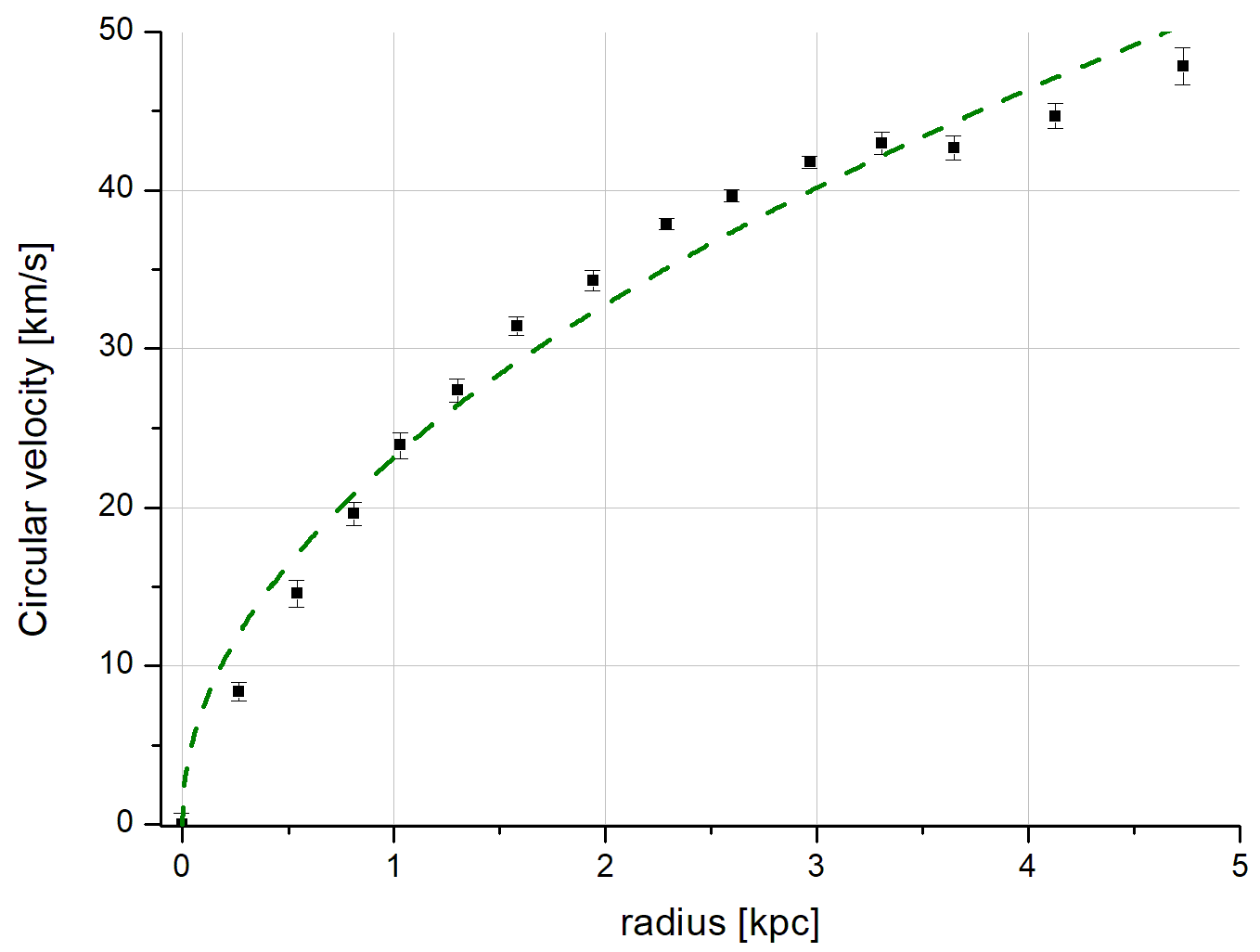}
	
	\caption{Squares with error bars are the synthetic universal rotation curve of dwarf galaxies modeled by E. V Karukes and P. Salucci \protect\cite{karukes_universal_2017}. The dashed green curve is a fitting of experimental data with equation (\protect\ref{veloce4}).}
	\label{fitkaruk}
\end{figure}

Karukes \& Salucci \cite{karukes_universal_2017} analyze a sample of dwarf disc galaxies to extend the Universal Rotation Curve (URC) framework to low-mass systems. 
They show that, once normalized by optical radius and velocity, the rotation curves exhibit a remarkably universal shape despite large variations in luminosity. 
Their mass modeling indicates that dwarf discs are strongly dark-matter dominated and are best described by cored halo profiles. 
They find that the core radius of the dark matter halo scales with the stellar disc scale length, pointing to a tight coupling between baryons and dark matter. 
On figure \ref{fitkaruk}, the rotation curve appears to increase monotonically.\\
\indent Within QMI, when the proper force acting on a system becomes smaller than $ma_{\rm min}$, Eq.~(\ref{PFD}) predicts that the acceleration saturates at the acceleration minimal bound. Astrophysical systems that are only weakly gravitationally bound therefore provide the most favorable environments for probing this low-acceleration regime.
In particular, owing to their comparatively small masses, dwarf galaxies can exhibit centripetal accelerations significantly below those encountered in massive galaxies. To first approximation, the Newtonian contribution may then remain minor over a large fraction of the rotation curve, allowing the dynamics to be effectively governed by the minimal-acceleration term. Equation (\ref{veloce3}) is then reduced to

\begin{equation} \label{veloce4}
	\boldsymbol{v}=\sqrt{ r a_{\rm min}}\ \hat{\boldsymbol{r}},
\end{equation}

\noindent and provides the minimal velocity rotation curve. The fitting of the synthetic data by equation (\ref{veloce4}) provides an estimation of the minimal acceleration 
\begin{equation} \label{aminvalue}
	a_{\rm min} = (1.75 \pm 0.06) \times 10^{-11} \text{m s}^{-2}.
\end{equation}

The fitting of the dwarf universal rotation curve with $M=0$ corroborates the recent claim that the stellar dynamics of dwarf galaxies are completely dominated by the putative dark matter halo \cite{battaglia_stellar_2022}.
This estimate nevertheless relies on the assumption that the Newtonian gravitational contribution remains below the minimal-acceleration scale everywhere in the dwarf galaxy. Depending on the mass distribution of the galaxy, the Newtonian force may become comparable to $ma_{\rm min}$ in some regions, particularly toward the central parts where the gravitational field is expected to be stronger than in the outskirts. As a result, moderate deviations from the pure minimal-acceleration regime may arise. Despite these limitations, the value of $a_{\rm min}$ inferred from \cite{karukes_universal_2017} still provides a useful estimate of the expected order of magnitude of the minimal acceleration scale. Other astrophysical tracers, however, may allow a more precise determination of $a_{\rm min}$.

\subsection{Milky Way rotation curve in Quantum Modified Inertia}
\indent Owing to the unprecedented quality of the third data release of the Gaia mission (Gaia DR3) \cite{vallenari_gaia_2023}, several recent analyses have suggested that the Milky Way (MW) rotation curve exhibits a Keplerian decline between Galactocentric radii of approximately 15 kpc and 27 kpc \cite{jiao_detection_2023, wang_mapping_2023}. Such a decrease in the circular velocity strongly disfavors a flat rotation curve.\\
\indent Recently, \cite{sofue_inner_2025} present an updated and high-precision MW rotation curve by combining HI, CO, maser and Gaia astrometric data. They construct a unified rotation curve extending from the Galactic Center to the outer halo, with improved resolution in both the inner bulge region and the outer disk. Their decomposition uses dynamical modeling of three components - bulge, disk, and dark halo without imposing photometric stellar-mass constraints or stellar-to-halo mass ratios. The resulting bulge and disk contributions are tightly constrained in the inner Galaxy, while the halo model is fit to the outer, nearly flat portion of the curve. The analysis indicates a relatively modest dark-matter halo, yielding a total Milky Way mass lower than many modern Gaia-based estimates (Table \ref{tab:sofue2025params}).

\begin{table}[h!]
	\centering
	\caption{Galactic mass model parameters from Sofue \& Kohno (2025). The third column shows the masses used in the modeling of figure \ref{milkyway}.}
	\begin{tabular}{lccc}
		\hline
		\textbf{Element}  & \textbf{Scale radius} & \textbf{Obs. Mass} & \textbf{QMI Mass} \\
		& \textbf{(pc)} & \textbf{($10^{11}\,M_{\odot}$)} & \textbf{($10^{11}\,M_{\odot}$)}\\
		\hline
		Bulge & $332.8 \pm 3.7$ & $0.127 \pm 0.002$ & 0.127 \\
		Disk & $5624.8 \pm 46.2$ & $1.352 \pm 0.011$ & 1.55 \\
		Halo & $22379.1 \pm 684.2$ & $0.599 \pm 0.019$ & 0  \\
		\hline
		\multicolumn{4}{l}{\footnotesize Note: Halo fit uses an NFW profile.}\\
	\hline
\end{tabular}
\label{tab:sofue2025params}
\end{table}
The authors also find indications that the rotation curve slightly declines at large radii, which contributes to a reduced inferred halo mass. Their curve exhibits strong internal consistency across tracers, producing one of the smoothest global Galactic rotation profiles to date.

\begin{figure}[h!]
	\center
	\includegraphics[width=9cm]{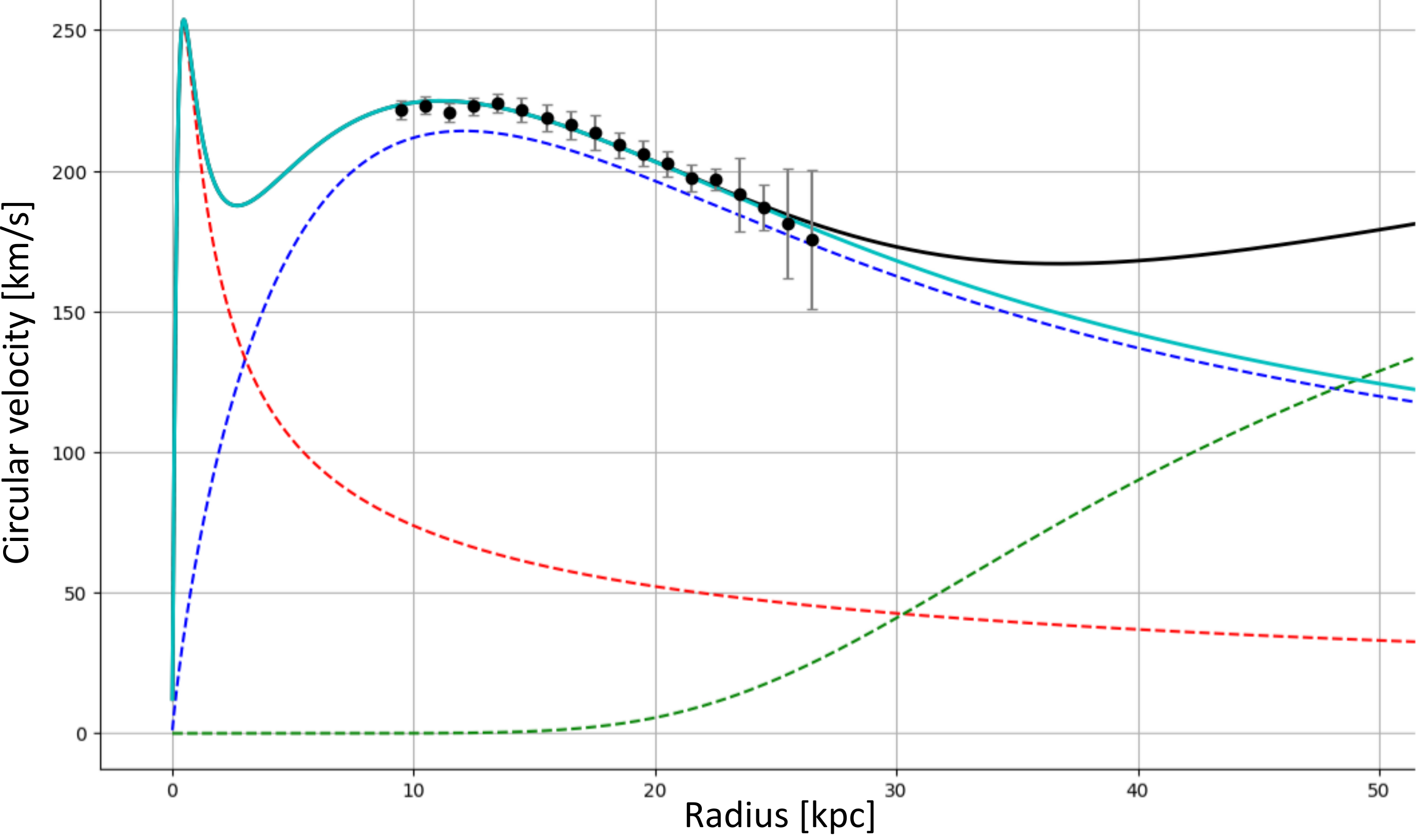}
	\caption{Plot of the velocity curve as function of the distance from the Center. Data from \protect\cite{jiao_detection_2023} is represented by circles with error bars. The dashed red, blue and green curves are the contributions to velocity respectively from the bulge, the disk (gas and stars) and the QMI minimal acceleration. The baryonic contents are listed in table \ref{tab:sofue2025params}.}
	\label{milkyway}
\end{figure}

In order to perform a simulation at very large radii, where the QMI minimal acceleration is expected to become dominant, it is more realistic to use an exponential disk \cite{freeman_disks_1970} rather than a Plummer profile \cite{plummer_problem_1911}. In practice, the scale lengths and masses of baryonic contents are conserved in the QMI modeling, except an increase of the disk mass of approximately $12\%$ when passing from a Plummer profile to an exponential disc. This is expected because the decomposition of \cite{sofue_inner_2025} was done under the spherical assumption, which gives a reasonable approximation to the bulge and the halo, while it underestimates the mass of the disc by $\sim 10$ percent compared to the value calculated for~a~thin~disc~\cite{sofue_rotation_2020}.

\indent The figure \ref{milkyway} shows that QMI is able to reproduce the putative MW Keplerian decline. According to the theory, the minimal acceleration effect is negligible at the Galactic Center, where the gravitational acceleration is stronger. As the gravitational acceleration decreases with distance, the hyperbolic function gradually increases the QMI minimal acceleration term. At $\sim 30$~kpc, the Newtonian velocity curve (in cyan) and the QMI velocity curve (in black) begin to split. The QMI predicts the circular velocity flattens before increasing smoothly. On the outskirts, the model also predicts that the speed is independent of the galaxy mass, because it is almost entirely governed by $a_{\rm min}$. One can see that the model is strongly dependent on the value of $a_{\rm min}$ at long distance and of the mass of the galaxy at short distance. For the value of $a_{\rm min}$ previsouly extracted from the universal rotation curve of dwarf galaxies, the model predicts that the MW rotation curve decreases at least until 35~-~40~kpc before rising again.\\
\indent Nevertheless, observational data currently extend only to 27 kpc, and probing the velocity curve at larger radii requires identifying tracers orbiting farther out. The Gaia mission provides a promising avenue in this respect, as it enables the detection and characterization of ultra-faint Milky Way satellites (UFMWs) at large distances \cite{simon_gaia_2018, yang_gaia_2023}.
An inspection of the orbital parameters reported in \cite{simon_gaia_2018} shows that UFMWs exhibit high velocities ($\sim$ 200-400 km/s), in several cases exceeding the maximum velocity measured before the onset of the Keplerian decline (224 km/s), qualitatively consistent with the model. However, these UFMWs preliminary estimates must be interpreted with caution due to the large uncertainties affecting their orbital parameters and the typically high orbital eccentricities of these systems.\\
\indent In addition, observational interpretation is complicated by the doubt of dynamical equilibrium in the Galactic disk due to past gravitational interactions, especially with the Sagittarius dwarf galaxy \cite{vasiliev_tango_2021}. Independent dynamical tracers, such as tidal streams, sometimes yield gravitational fields at odds with those inferred from simple rotation models. Since a better confirmation of the Keplerian decline is needed, the QMI success in plotting the MW rotation curve should be seen as a consistency check rather than a strong evidence in support of the theory. 

\subsection{DDO 52 rotation curve in Quantum Modified Inertia}

\begin{figure}[h!]
	\center
	\includegraphics[width=9cm]{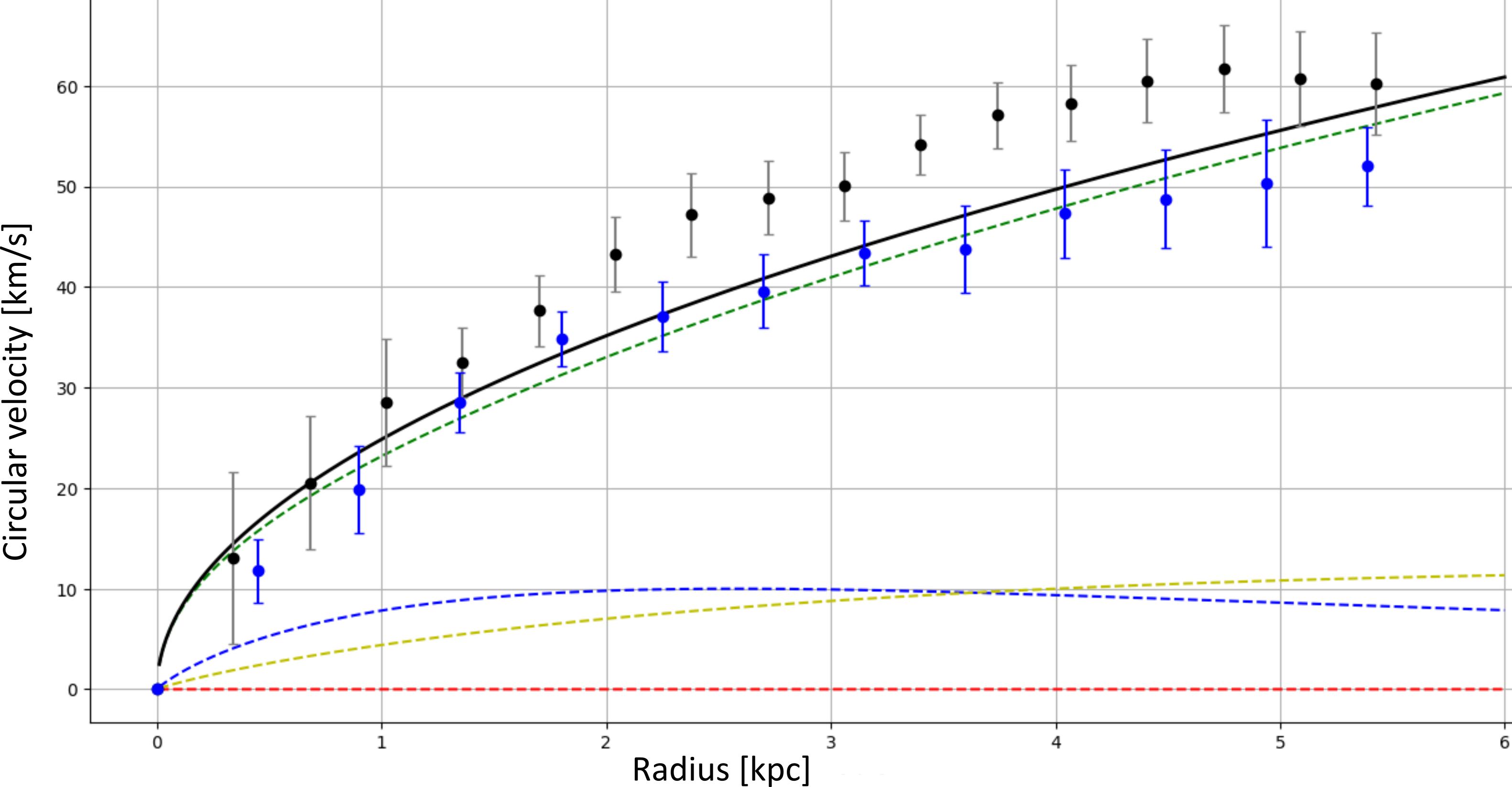}
	\caption{Rotation curve of DDO52. The two dataset are extracted from \cite{oh_high-resolution_2015} (black dots) and \cite{iorio_little_2017} (blue dots). The plain black line is the modeled rotation curve.The dashed red, blue, yellow and green curves are the contributions to velocity respectively from the bulge, the stellar disk, the gas disk and the QMI minimal acceleration. The baryonic contents are listed in table \ref{tab:DDO52Oh}.}
	\label{DDO52fig}
\end{figure}

\indent QMI indicates that the minimal acceleration term is dominant in low gravitational fields, therefore dwarf galaxies provide a natural laboratory for observing this minimal acceleration effect.
The rotation curve of the isolated dwarf DDO52 is provided by LITTLE THINGS \cite{oh_high-resolution_2015} and later by a refining of the galaxy inclination \cite{iorio_little_2017} studies (figure \ref{DDO52fig}). The simulation takes in account the stellar disk and the HI disk, both represented by Freeman exponential disks. The masses are taken from \cite{oh_high-resolution_2015} and the scale radii are adjusted to match the velocity contributions depicted in \cite{mancera_pina_galaxy-halo_2025}. The halo mass is barely estimated because the rotation curve is still rising at the last measured points. The decompositions of DDO 52 are reported in the table \ref{tab:DDO52Oh}, for both LITLLE THINGS and QMI.

\begin{table} [h!]
\centering
\caption{Stellar, gas and halo parameters of DDO 52. The first column are scale radius adjusted to recover the contributions of gas and stellar contents plotted in \cite{mancera_pina_galaxy-halo_2025}. The second column are observed disk masses extracted from from \cite{oh_high-resolution_2015}. The third column are the masses used in the QMI model to plot the rotation curve in figure \ref{DDO52fig}.}
\begin{tabular}{lccc}
	\hline
	\textbf{Element}  & \textbf{Scale radius} & \textbf{Obs. Mass} & \textbf{QMI Mass} \\
	& \textbf{(kpc)} & \textbf{($10^{8}\,M_{\odot}$)} & \textbf{($10^{8}\,M_{\odot}$)}\\ \hline
	Gas & $4$ & $ 3.343 $ & $3.343$ \\
	Stars & $1.2$ & $0.7$ & $0.7$ \\
	Halo & $1.33$ & $> 10$ & $0$  \\ \hline
	\multicolumn{4}{l}{\footnotesize Note: Halo fit uses a pseudo-isothermal profile.}\\
	\hline
\end{tabular}
\label{tab:DDO52Oh}
\end{table}

The two datasets illustrate the diversity of data and the impact of a slight misestimation of galaxy disk inclination. 
With the value of $a_{\rm min} = 1.8\times 10^{-11} \text{m s}^{-2}$, the agreement is excellent although the model is simplified to two thin exponential disks. Inasmuch the gravitational acceleration due to baryonic content is weak, the circular velocity is almost entirely due to the QMI minimal acceleration, largely overcoming the choice of the baryonic profile model. 

\subsection{Radial acceleration relation in QMI}

\begin{figure}[h!]
\centering
\includegraphics[width=9cm]{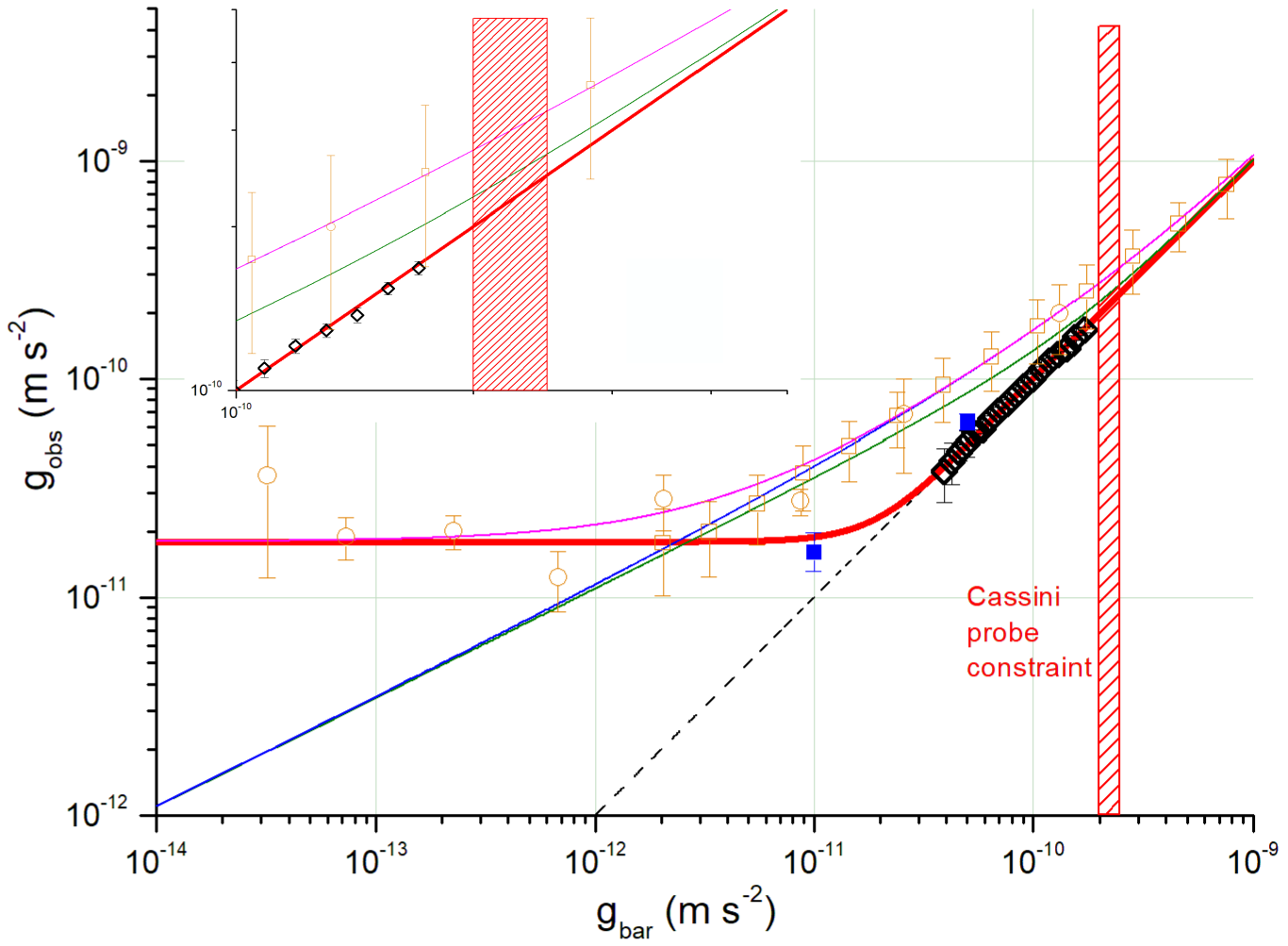}

\caption{Plot of the RAR representing the observed acceleration $g_{\rm obs}$ as a function of the acceleration due to baryons $g_{\rm bar}$. The dark diamond is the MW acceleration data calculated from \cite{jiao_detection_2023}. The orange square and circle are respectively the binned data from SPARC for LTG and ETG, and the SPARC binned data for dwarf galaxies \cite{lelli_one_2017, lelli_sparc_2016}. Blue squares are the observed accelerations of ultra-wide binaries \cite{yoon_probing_2025}. The black dashed line is the Newtonian acceleration $g_{\rm obs}=g_{\rm bar}$. Three MOND IF are plotted : $g_{\rm obs}^{\rm St}$ in green, $g_{\rm obs}^{\rm SPARC}$ in blue and $g_{\rm obs}^{\rm SPARC+f}$ in purple. The Cassini constraint \cite{desmond_tension_2024} is represented by a red hatched rectangle. The red curve is the QMI RAR $g_{\rm obs}^{\rm QMI}$. The inset is a zoom of the RAR near the Cassini constraint range.}
\label{RAR}
\end{figure}

In order to extend and summarize these galaxy rotation curves, we plot a RAR, which spans over Newtonian and low acceleration regimes. In the framework of MOND, the RAR emerges as a natural consequence of the modification of dynamics at low accelerations. MOND predicts a tight correlation between the observed centripetal acceleration and that expected from the baryonic mass distribution, with a characteristic transition around the acceleration scale $a_0 \simeq 1.2\times10^{-10}\,\mathrm{m\,s^{-2}}$. This relation, observed across a wide range of galaxy types and masses, is often regarded as one of the strongest empirical successes of MOND.
Figure~\ref{RAR}, reproduced from \cite{lelli_one_2017}, illustrates this relation through the SPARC data \cite{lelli_sparc_2016}. 
In addition to the observed deviation from Newtonian predictions, the authors of \cite{lelli_one_2017} suggest the presence of a ''flattening'' of the acceleration at a value close to $9\times10^{-12}\,\mathrm{m\,s^{-2}}$, remarkably near the adopted $a_{\min} = 1.8\times10^{-11}\,\mathrm{m\,s^{-2}}$. In Fig.~\ref{RAR}, dwarf galaxy data are binned without imposing tidal quality cuts, which accounts for the  differences relative to \citet{lelli_one_2017}.
The general behavior of the RAR is well captured by the present model, as shown by the red curve in Figure~\ref{RAR}. 
Notably, this curve is not a fit, but a direct plot of the expected acceleration from equation~(\ref{PFD}) using the value $a_{\min} = 1.8\times10^{-11}\,\mathrm{m\,s^{-2}}$ derived from the data of \cite{karukes_universal_2017}. In addition, the following MOND interpolations functions (IF) plotted on figure \ref{RAR} are:

\begin{eqnarray} \label{RAR_IF}
&\text{SPARC}& g_{\rm obs}^{\rm SPARC} = \frac{g_{\rm bar}}{1-\text{exp}(\sqrt{(g_{\rm bar}/g_\dagger)})} \nonumber \\
&\text{SPARC+flat.}& g_{\rm obs}^{\rm SPARC+f} = g_{\rm obs}^{\rm SPARC} + \hat{g}~ \text{exp}\left(-\sqrt{g_{\rm bar}g_\dagger/\hat{g}^2}\right) \nonumber  \\
&\text{Standard}& g_{\rm obs}^{\rm St} = \sqrt{\frac{g_{\rm bar}^2}{2} +\sqrt{\frac{g_{\rm bar}^4}{4}+g_{\rm bar}^2 a_0^2}},\\
\end{eqnarray} 

with $\hat{g} \simeq 1.2 \times 10^{-10}$ m s$^{-2}$ and $g_\dagger \simeq 9.2 \times 10^{-12}$ m s$^{-2} $ \cite{lelli_one_2017}. In this MOND interpretation of the flattening, $\hat{g}$ and $g_\dagger$ are respectively equivalent to $a_0$ and $a_{\rm min}$.

Still in figure \ref{RAR}, the QMI RAR is given by:
\begin{equation} \label{QMI RAR}
g_{\rm obs}^{\rm QMI} = g_{\rm bar} + a_{\rm min}\left[1-\text{tanh}\left(\frac{g_{\rm bar}}{a_{\rm min}}\right) \right].
\end{equation}
In contrast, $g_{\rm obs}^{\rm QMI}$ does not constitute an arbitrary interpolation function, since its functional form arises directly from the relativistic framework underlying the model. In addition, the hyperbolic tangent leads to a sharper transition between the low-acceleration and Newtonian regimes than the interpolation functions commonly employed in MOND.

While both the Newtonian and low-acceleration regimes appear broadly consistent with the SPARC data \cite{lelli_sparc_2016}, the transition region is less accurately reproduced. It should however be emphasized that the SPARC analysis relies significantly on assumptions regarding the stellar mass-to-light ratio, and that the inferred shape of the radial acceleration relation (RAR) depends sensitively on the adopted value of the MOND acceleration scale $a_0$. Current estimates still involve uncertainties at the $\sim 20$ \% level \cite{lelli_one_2017}. In addition, recent studies suggest that the stellar mass-to-light ratio may not be uniform across galactic disks and could depend on variations of the initial mass function (IMF) \cite{dabringhausen_integrated_2023}. Such effects may influence the inferred baryonic distribution and therefore modify the detailed structure of the RAR, particularly in the transition regime.\\
\indent Using the Milky Way rotation curve data from \citet{jiao_detection_2023} together with the Newtonian rotation curve shown in Fig.~\ref{milkyway} (cyan curve), the Milky Way velocity profile is recast into a radial acceleration relation (RAR), shown in Fig.~\ref{RAR}. The inferred baryonic acceleration 
$g_{\rm bar}$ depends on the adopted decomposition of the Galactic mass components. Such variations, however, do not displace the Milky Way data points away from the Newtonian slope, but merely shift them along it. It should nevertheless be stressed that the Milky Way data points are plotted under the assumption that the observed decline is genuinely Keplerian, an interpretation that must be treated with caution if the Galactic disk is not in dynamical equilibrium.\\
\indent On the other hand, the Cassini spacecraft provided highly precise measurements in the vicinity of the Saturnian system, offering stringent constraints on possible deviations from Newtonian gravity. Within the MOND framework, the external field effect (EFE) generally induces a quadrupolar correction to the gravitational potential. Analyses of the Cassini tracking data indicate that the amplitude of such a quadrupole is tightly constrained, leading to a tension with some MOND formulations \cite{desmond_tension_2024}. Consequently, Solar-System observations in the acceleration range ${g_{\rm bar} \sim (2 - 2.5) \times10^{-10}}$~m~s$^{-2}$ provide a useful benchmark for testing modified-dynamics scenarios involving external-field effects. The RAR given by the QMI is however in accordance with the Cassini trajectory data, as shown on the inset of figure \ref{RAR}.\\
\indent Since QMI does not explicitly incorporate an external-field effect analogous to that of MOND, no additional quadrupolar correction is expected to arise naturally within the Solar System. In the present framework, the gravitational field generated by the Sun reaches the minimal-acceleration scale only near the inner boundary of the Oort cloud. Nevertheless, several observational studies suggest that the dynamics of low-acceleration subsystems embedded within the gravitational field of a larger structure may be compatible with MOND-like external-field effects. The extent to which similar environmental or tidal effects could emerge effectively within QMI therefore deserves further investigation.

\subsection{Wide binaries}
Far below the galactic scale, wide binaries separated by ${\sim 1-100}$~kAU are sensitive probes of gravity in the low-acceleration regime, as a fonction of their masses. 
In Newtonian dynamics, their orbital velocities decrease with separation following Kepler's law, whereas MOND predicts a flattening or excess velocity when the mutual acceleration drops below the MOND scale $a_0 \simeq 1.2 \times 10^{-10}\,\mathrm{m\,s^{-2}}$. 
While wide binaries provide a potential low-acceleration test of MOND, the observational results remain debated. Some studies using \textit{Gaia} DR2/DR3 report relative velocities slightly higher than predicted by Newtonian gravity, which could hint at MOND effects \cite{chae_measurements_2024}. However, other analyses find that the data are consistent with Newtonian expectations once uncertainties, projection effects, contamination from chance alignments, and orbital eccentricities are properly accounted for, \cite{banik_strong_2023}.\\
\indent In QMI, the situation is more complex. The theory predicts that orbital parameters are affected by the minimal acceleration, but in the low $10^{-11}\,\mathrm{m\,s^{-2}}$ range, due to the sharper transition between Newtonian and minimal acceleration regime (figure \ref{RAR}). The selected wide binaries must therefore have lower mass or greater distances than many of the ones chosen in recent studies. For example, for a wide binary composed of a solar mass star and a red dwarf star, the expected distance to reach the minimal acceleration regime is close to $\sim 17000$~AU, and has to be increased for even more massive stars. The minimal acceleration could consequently only be observed in ultra weakly bonded binaries.\\
\indent Using Gaia DR3 wide binaries at separations up to $\sim50$~kAU, low-acceleration gravity tests with ultra-wide binaries were recently reassessed while carefully accounting for perspective effects in proper motions \cite{yoon_probing_2025}. Even after these corrections, the observed acceleration excess over Newtonian predictions persists, though improved radial velocity data are still required.
In particular, for ${g_{\rm bar}=1\times10^{-11}}$~m~s$^{-2}$ and $g_{\rm bar}=5\times10^{-11}$ m s$^{-2}$, the inferred observed accelerations are respectively $g_{\rm obs}=1.61^{+0.37}_{-0.29}\times10^{-11}\,\mathrm{m\,s^{-2}}$ and $g_{\rm obs}=6.31^{+0.6}_{-0.5}\times10^{-11}\,\mathrm{m\,s^{-2}}$. These two results from \cite{yoon_probing_2025} are reported in the RAR figure \ref{RAR} and seem to suggest a better agreement for QMI than for MOND, although the uncertainty on the data is important and the systematic effects are still a matter of debate. Actually, in the considered acceleration range, the modified acceleration boost is still small. The next release of Gaia data (DR4) could possibly extend the acceleration range. Ultra-weakly bonded binaries, in the $10^{-12}\,\mathrm{m\,s^{-2}}$ range, would constitute a stronger test for the theory.\\

\subsection{Tully-Fisher relation in QMI}
The TFR \cite{tully_new_1977} is an empirical correlation observed in spiral galaxies between their luminosity (or baryonic mass) and the rotation velocity measured in the flat part of the rotation curve. 
Specifically, more luminous spiral galaxies exhibit higher rotational velocities, implying a strong link between the total baryonic content and the depth of the gravitational potential. 

Brightness is proportional to $v^\alpha$, with $\alpha$ close to 4, depending on the observed wavelength \cite{torres-flores_ghasp_2011}.
In the infrared domain, the total luminosity of the galaxy is approximately proportional to the mass of stars. Therefore, we deduced that the stellar mass is proportional to $v^4$. MOND remarkably predicts this relationship by considering $v$ as the speed limit at large $r$ \cite{mcgaugh_baryonic_2012}, although, from an observational point of view, the measured velocity is not necessarily the velocity at high radius $r$ but rather the maximum velocity. The MOND TFR arises naturally as a direct consequence of the modified force law in the low-acceleration regime:
\begin{equation} \label{Tully_MOND}
M = \frac{v^4}{G a_{0}}
\end{equation}

In the QMI framework, the asymptotic behavior differs from the standard MOND picture. At first sight, Eq.~(\ref{veloce4}) suggests that the orbital velocity does not converge toward a strictly constant value at large radii, since the minimal-acceleration regime implies \(v \propto \sqrt{r}\). The emergence of a Tully-Fisher-like relation is therefore less direct than in MOND and requires a more careful interpretation.\\
\indent In Newtonian dynamics, the gravitational acceleration generated by an enclosed baryonic mass \(M_{\rm bar}\) is

\begin{equation}
	a_N=\frac{G M_{\rm bar}}{r^2}.
\end{equation}
By contrast, in the QMI low-acceleration regime, the effective acceleration asymptotically approaches

\begin{equation}
	a_{\rm QMI}\simeq a_{\rm min}.
\end{equation}
If the observed orbital motion is nevertheless interpreted within a Newtonian framework, the enclosed dynamical mass inferred from the kinematics becomes

\begin{equation}\label{min-mass}
	M_{\rm dyn}=\frac{a_{\rm min} r^2}{G}.
\end{equation}
This quantity should not be interpreted as a genuine additional baryonic mass, but rather as an effective dynamical mass associated with the modified inertial regime. In this sense, the minimal-acceleration sector of QMI mimics the presence of an increasing enclosed mass when analyzed through standard Newtonian dynamics.

Assuming circular motion, \(a=v^2/r\), the asymptotic velocity in the minimal-acceleration regime satisfies

\begin{equation}
	v=\sqrt{r a_{\rm min}}.
\end{equation}
Substituting this relation into Eq.~(\ref{min-mass}) yields

\begin{equation}\label{Tully-QMI}
	M_{\rm dyn}=\frac{v^4}{G a_{\rm min}}.
\end{equation}
The model therefore reproduces a Tully-Fisher-like scaling relation between the inferred dynamical mass and the asymptotic orbital velocity. Unlike MOND, however, this relation does not emerge from strictly flat rotation curves, but from the effective dynamical mass associated with the minimal-acceleration regime.
In addition, Eqs.~(\ref{Tully-QMI}) and (\ref{Tully_MOND}) differ through the characteristic accelerations entering the normalization, namely \(a_{\rm min}\) for QMI and \(a_0\) for MOND. Expressed in logarithmic form, using solar masses and velocities in km\,s\(^{-1}\), the corresponding relations become

\begin{eqnarray}\label{Tully_BOTH}
	\mathrm{MOND}&:	\mathrm{log} (M) = 4 \mathrm{log} (v) + 1.80, \\
	\mathrm{QMI}&:	\mathrm{log} (M) =  4 \mathrm{log} (v) + \mathbf{2.62}.
\end{eqnarray}
Although both frameworks predict the same asymptotic slope, they lead to different zero-point offsets (ZPOs). The observational determination of the ZPO therefore provides, in principle, a possible discriminant between the two approaches. For instance, Ref.~\cite{gogate_budhies_2023} reported \(\mathrm{ZPO}=2.80\pm0.36\) for galaxies in the Ursa Major supercluster, while Ref.~\cite{sharma_tully-fisher_2024} obtained \(\mathrm{ZPO}=3.34\pm0.53\) for the stellar Tully--Fisher relation and \(\mathrm{ZPO}=3.16\pm0.61\) for the baryonic relation.
Nevertheless, the observational literature on the Tully-Fisher relation remains broad and method-dependent, with the inferred zero-point offset being sensitive to sample selection, distance calibration, stellar mass-to-light ratios, and the adopted fitting procedure. A more systematic analysis would therefore be required before drawing robust conclusions regarding the relative consistency of QMI with current Tully-Fisher data.

\section{Discussion}\label{sec3}

\subsection{General relativity statement}
\emph{Towards the strong-field regime}

The present formulation of QMI is developed starting from a special-relativistic framework and is therefore not expected to provide a complete description of strong gravitational fields. A consistent generalisation to curved spacetime, fully compatible with general relativity in the appropriate limit, remains to be constructed.
In the weak-field regime, QMI introduces a maximal proper acceleration scale of the form
\[
a_{\rm max} \sim \frac{2 m c^3}{\hbar},
\]
which depends on the rest mass of the test particle. For elementary particles, this scale can be extremely large: for instance, \(a_{\rm max} \sim 10^{29}\,\mathrm{m\,s^{-2}}\) for electrons and \(a_{\rm max} \sim 10^{32}\,\mathrm{m\,s^{-2}}\) for protons. In comparison, the corresponding Planck acceleration is significantly higher, so that such bounds remain well below the regime where quantum-gravitational effects are expected to dominate.
It is therefore natural to expect that, in the presence of strong gravitational fields, the QMI framework should recover standard general relativistic dynamics in addition to possible corrections associated with acceleration bounds. 
A consistent extension of QMI to curved spacetime may nevertheless provide insights into the interplay between quantum kinematical bounds and strong gravitational fields, potentially including regimes close to or inside horizons.\\

\noindent \emph{F=0 case}\\
\indent In the present formulation, the minimal acceleration contribution is defined as a vector aligned with the applied force. In the absence of an external force, however, no preferred spatial direction is selected, and the definition of a unique acceleration vector becomes ambiguous. This issue is not specific to QMI and appears in other modified inertia or modified gravity frameworks, where nonlinear behavior at low accelerations raises questions about the treatment of vanishing forces. For example, this question is also present in MOND, depending on which interpolation function is used. The SPARC RAR IF $g_{\rm obs}^{\rm SPARC+f}$ presents an undefined solution when $g_{\rm bar}= 0$.\\
\indent A generalized QMI should provide the appropriate covariant dynamics that could presumably resolve the ambiguities encountered in non-generalized formulations, particularly regarding directionality and inertial motion. In this sense, a minimal acceleration could be meaningfully defined as a geometry-dependent bound rather than as a universal dynamical constant.\\
\indent However, within the scope of the present work, the phenomenological applications always involve configurations in which an external field provides a well-defined direction for the acceleration, accordingly, there is no practical consequence.\\


\subsection{A dependence to the redshift?}
However, although a generalized QMI is not developed yet, we can still draft a cosmological consequence of the minimal acceleration.  Here, $a_{\rm min}$ is presented as a universal constant, in the sense that its value should be everywhere the same because of the isotropy of space and the homogeneity of the observable Universe at a great scale. However, does $a_{\rm min}$ remain constant over cosmic time ? Under the conjecture (\ref{conjecturette-tau-}), the minimal acceleration value directly depends on the radius of the causality sphere, this could mean that its value decreases with time and, conversely, that $a_{\rm min}$ was greater in the past.\\
\indent Let us consider the scale factors $a_S(t_0)$ and $a_S(t)$, respectively at the present epoch $t_0$ and at any other epoch $t$, as a function of the redshift $z$:
\begin{equation} \label{redshift}
	\frac{a_S(t_0)}{a_S(t)}=1+z.
\end{equation}
We can thus define the radius of the causality sphere at any time with:
\begin{equation} \label{radius}
	\frac{R_u(t)}{R_u(t_0)}=\frac{a_S(t)}{a_S(t_0)}.
\end{equation}
Using equations (\ref{redshift}) and (\ref{radius}), we can express the redshift as a function of $R_u(t_0)$ and $R_u(t)$:
\begin{equation} \label{zradius1}
	z=\frac{R_u(t_0)}{R_u(t)}-1.
\end{equation}
Moreover, the value of $a_{\rm min}$ theoretically depends on the radius of the causality sphere:
\begin{equation} \label{zradius2}
	a_{\rm min}(t)=\frac{c^2}{\pi R_u(t)},
\end{equation}
so we can express $a_{\rm min}(t)$ as a function of the redshift:
\begin{equation} \label{zradius3}
	a_{\rm min}(t)=\frac{c^2(1+z)}{\pi R_u(t_0)}.
\end{equation}
Consequently, it appears that $a_{\rm min}$ could vary linearly with the redshift $z$ and that $a_{\rm min}$ could have been greater during the early Universe. The Magneticum collaboration \cite{mayer_lambdacdm_2022} depicts the evolution of the MOND acceleration scale $a_0$ with $z$ and observe a constant decrease of $a_0$ with $z$. Although the $a_0$ MOND scale factor and $a_{\rm min}$ are conceptually different, it can leave a hint concerning the evolution of $a_{\rm min}$ with the redshift. Furthermore, the formation of early galaxy discs could also reveal some different features in the presence of a stronger minimal acceleration. This dependence with $z$ however depends on the validity of the conjecture establishing the expression for the minimum acceleration.\\

\subsection{Universality of the minimal acceleration ?}

\indent Finally, the application examples of the minimal acceleration are all related to galaxy rotation curves and thus to gravity. This  could suggest that the quantum bound on acceleration exclusively concerns gravity. In fact, regarding the maximal acceleration bound, several applications proposed in the literature are not only related to gravity but also not exhaustively to the spin of particles \cite{papini_spin_2017}, gravitation and electric field \cite{gasperini_kinematic_1991}, and corrections of the Lamb shift \cite{lambiase_maximal_1998}. In QMI, as the minimal and maximal acceleration bounds are unified in the same framework, it seems that there is no justification to restrict the minimal acceleration to gravity. Both quantum bounds on acceleration presumably apply to all possible forces or interactions and should be considered as a general dynamical effect, independently of the nature of the force applied on the accelerating object. This approach promotes the previous development of modified inertia instead of solely searching for a modification of gravitation's law.

\section{Conclusions}

In summary, starting on SR we have proposed a novel dynamical principle in which the proper time of an accelerated system is assumed to be bounded between two extremal values. Within this framework, upper and lower bounds on the coordinate acceleration naturally emerge. It is further conjectured that these bounds can be related to quantum speed limits, providing a possible quantum-kinematical interpretation of the construction.\\
\indent Regarding consistency with established physical laws, when the acceleration slightly exceeds $a_{\rm min}$, this model rapidly recovers Einsteinian dynamics and reduces to the Newtonian law $a=F/m$ for low velocities. This principle of dynamics thus closely resembles standard dynamics, diverging only near the lower and upper quantum acceleration bounds, which are extremely small (about 12 orders of magnitude below Earth's gravity) and extremely high, depending on the object mass.\\
\indent The most evident context in which the minimal acceleration appears to dominate is the rotation curves of galaxies. Far from the galactic bulge, the acceleration due to Newtonian gravity is comparable to $a_{\rm min}$ or even smaller. Remarkably, this minimal acceleration effect has implications for the dark matter problem, which remains debated between proponents of cold dark matter to preserve standard dynamics and advocates of MOND, which modifies Newtonian dynamics by introducing a characteristic acceleration scale.\\
\indent The present formulation differs from MOND in that it introduces a minimal acceleration scale that is conceptually distinct from the MOND parameter $a_0$, even though both scales are of comparable magnitude. In the low-acceleration regime, the model generically leads to a gradual departure from strictly flat rotation curves, and predicts a slow radial evolution of the orbital velocity. Such behaviour can be confronted with observations of dwarf galaxies, wide binaries, and empirical scaling relations such as the baryonic Tully-Fisher relation. Furthermore, this theory differs from "ordinary" cold dark matter, which is assumed to be abundantly distributed in galaxy disks and outskirts to account for rotation curves. Although the theory does not claim to exclude cold dark matter or exotic particles such as axions, it implies that the amount of dark matter needed to explain celestial motions could be dramatically lower.\\
\indent The results point out that QMI is successful in reproducing several astrophysical observations in which deviations from the Newtonian dynamics are observed at low accelerations. These results should however be considered as consistency checks rather than a strong confirmation, due to uncertainties affecting the observations.\\
\indent Finally, a conjectured prediction is that the minimal acceleration was likely larger in the past, potentially exerting a significant influence on the formation of galaxies and other large-scale structures in the Universe.

\backmatter

\bmhead{Acknowledgements}

I would like to thank Olivier Bienaym\'e who patiently gave a lot of informations about galaxy rotation curves and  bibliography about dark matter. I express my gratitude to Fran\c cois Vernotte, Jos\'e Lages, David Viennot, Federico Lelli and the anonymous referee for their valuable comments or suggestions concerning the theoretical calculations or astrophysical observations.
Finally, I would like to extend my warmest thanks to Annie Robin for her constant support and insightful suggestions, which have helped to refine this study.

\section*{Declarations}

\begin{itemize}
\item The author declare no competing interest.
\item Data availability: this is a theoretical study and no experimental data has been generated.
\item Code availability: the code generated during the current study is available from the corresponding author on reasonable request.
\end{itemize}

\noindent

\bigskip

\begin{appendices}

\section{Quantum speed limit}
\label{appendixA}

\subsection{Levitin \& Margolus speed limit}
There are many different interpretations of the time-energy uncertainty relation,  depending on the chosen definition of time \cite{Busch08}. One is that a quantum state with energy dispersion ${\Delta E}$ takes a time
\begin{equation} \label{DeltaT}
	\Delta t \ge \frac{\hbar}{2 \Delta E},
\end{equation}
to evolve between two distinguishable orthogonal states, known as the Mandelstam-Tamm speed limit. However, it has been proposed that the evolution time of a system has a more stringent limit, depending on its average energy $E$ \cite{margolus_maximum_1998}, for which Levitin et Margolus found a time defining the maximum speed of orthogonality evolution. If at time $t=0$, a quantum system is described by a superposition of the energy eigenstates:
\begin{equation} \label{eigen0}
	|\Psi(0) \rangle = \sum_n c_n |E_n \rangle,
\end{equation}
then, at a time $t$,
\begin{equation} \label{eigent}
	|\Psi(t) \rangle = \sum_n c_n\text{e}^{- \frac{i E_n t}{\hbar}}|E_n \rangle.
\end{equation}
The smallest time between $|\Psi(0) \rangle$ and $|\Psi(t) \rangle$ is given by the orthogonality relation:
\begin{equation} \label{orthopsi}
	\langle \Psi(0)|\Psi(t) \rangle = \sum_n |c_n|^2\text{e}^{- \frac{i E_n t}{\hbar}}=0.
\end{equation}

By using the trigonometry relation $\text{cos}(x) \ge
1 - 2 \pi (x + \text{sin}(x))$, which is valid for $x \ge 0$, the real part is developped as follows:
\begin{eqnarray} \label{reorthopsi}
	\mathcal{R}\{\langle \Psi(0)|\Psi(t) \rangle\} &=& \sum_n |c_n|^2\text{cos}{ \frac{ E_n t}{\hbar}},\\
	&\ge& \sum_n |c_n|^2\left(1-\frac{2}{\pi}\left(\frac{ E_n  t}{\hbar}+\text{sin}\left(\frac{ E_n t}{\hbar}\right) \right)\right), \\
	&\ge& 1-\frac{2}{\pi}\left(\frac{ E t}{\hbar}+\mathcal{I}\{\langle \Psi(0)|\Psi(t) \rangle\}\right).
\end{eqnarray}
To equate the whole equation to 0, we have to equate with 0 both the real and imaginary parts of (\ref{orthopsi}), so we obtain
\begin{equation} \label{LVtime}
	1-\frac{2}{\pi}\frac{ E t}{\hbar}=0,
\end{equation}
and an expression of the minimal time of evolution, also called the Levitin and Margolus QSL, directly comes:
\begin{equation} \label{DeltaTp+}
	\tau_{QSL}^+ = \frac{\hbar \pi}{2 E},
\end{equation}
with the initial state energy defined to be zero. In this sense, the time of evolution between the two states of the system gives the frequency of an associated clock. 
Furthermore, if $\Delta E = E$, the Mandelstam-Tamm speed limit (\ref{DeltaT}) and Levitin-Margolus limit (\ref{DeltaTp+}) are equal. $\Delta E > E$ implies a smaller $\Delta t$ which is excluded by (\ref{DeltaTp+}), thus $\Delta E$ is restricted to values $\Delta E \le E$.\\
\indent The Mandelstam--Tamm bound (\ref{DeltaT}) and the Margolus--Levitin bound (\ref{DeltaTp+}) constitute the two most widely used formulations of the quantum speed limit (QSL). Generalizations of these bounds have also been proposed, including the Margolus-Levitin--Toffoli formulation, which retains a closely related structure \cite{levitin_fundamental_2009}. More generally, the theory of quantum speed limits has undergone significant developments in recent years, with extensions toward open quantum systems, semiclassical and classical regimes \cite{shanahan_quantum_2018, garcia-pintos_unifying_2022,bolonek-lason_classical_2021}.

\section{Mass models used for the baryonic components}
\label{app:massmodels}

In this work, the baryonic mass distribution of galaxies is modeled using a thin exponential disk following Freeman \citep{freeman_disks_1970} and a spherical bulge described by the Hernquist profile \citep{hernquist_analytical_1990}. These models are widely used in galactic dynamics as they provide simple analytic expressions for both the mass distribution and the corresponding circular velocity profiles.

\subsection{Exponential stellar disk (Freeman disk)}

The surface mass density of a thin exponential disk is given by
\begin{equation}
	\Sigma(R) = \Sigma_0 \exp\left(-\frac{R}{R_d}\right),
\end{equation}
where $\Sigma_0$ is the central surface density and $R_d$ is the disk scale length. The total disk mass is
\begin{equation}
	M_d = 2\pi \Sigma_0 R_d^2.
\end{equation}

The contribution of the disk to the circular velocity in the mid-plane is computed using the exact expression derived by \citet{freeman_disks_1970},
\begin{equation}
	v_d^2(R) = \frac{G M_d}{2 R_d} y^2 
	\left[ I_0(y) K_0(y) - I_1(y) K_1(y) \right],
\end{equation}
where $y = R / (2 R_d)$ and $I_n$ and $K_n$ are the modified Bessel functions of the first and second kind, respectively. This expression is evaluated numerically in the Python implementation using standard scientific libraries.

\subsection{Spherical bulge (Hernquist profile)}

The bulge component is modeled using the Hernquist density profile,
\begin{equation}
	\rho_b(r) = \frac{M_b}{2\pi} \frac{a}{r (r+a)^3},
\end{equation}
where $M_b$ is the total bulge mass and $a$ is the bulge scale radius. The enclosed mass within radius $r$ is
\begin{equation}
	M_b(r) = M_b \frac{r^2}{(r+a)^2}.
\end{equation}

The corresponding circular velocity profile is obtained from
\begin{equation}
	v_b^2(r) = \frac{G M_b r}{(r+a)^2}.
\end{equation}

This analytic form makes the Hernquist profile particularly convenient for numerical implementations and ensures a finite total mass and a realistic central cusp.

\subsection{Total baryonic contribution}

The total baryonic circular velocity is computed by adding the disk and bulge contributions in quadrature,
\begin{equation}
	v_{\mathrm{bar}}^2(R) = v_d^2(R) + v_b^2(R),
\end{equation}
assuming axisymmetry and negligible cross terms. This baryonic rotation curve is then used as input for the dynamical model explored in the main text.

\subsection{Numerical implementation}

All expressions presented above are implemented in Python using standard numerical libraries. The modified Bessel functions appearing in the disk contribution are evaluated using \texttt{scipy.special}. 
This modeling framework provides a flexible and physically motivated description of the baryonic mass distribution, suitable for comparison with observed galaxy rotation curves.

\end{appendices}


\bibliography{biblio2} 

\end{document}